# Room-temperature detection of single 20 nm super-paramagnetic nanoparticles with an imaging magnetometer

*Michael Gould*[*a], *Russell Barbour*[b], *Nicole Thomas*[a], *Hamed Arami*[c], *Kannan M. Krishnan*[c], *Kai-Mei Fu*[a,b]

[a]Dept. of Electrical Engineering, Univ. of Washington, Box 352500, Seattle, WA, USA 98195-2500; [b]Dept. of Physics, Univ. of Washington, Box 351560, Seattle, WA, USA 98195-1560; [c]Dept. of Material Science and Engineering, Univ. of Washington, Box 352120, Seattle, WA, USA 98195-2120

Abstract: We demonstrate room temperature detection of single 20 nm super-paramagnetic nanoparticles (SPNs) with a wide-field optical microscope platform suitable for biological integration. The particles are made of magnetite ($Fe_3O_4$) and are thus non-toxic and biocompatible. Detection is accomplished via optically detected magnetic resonance imaging using nitrogen-vacancy defect centers in diamond, resulting in a DC magnetic field detection limit of 2.3 $\mu$T. This marks a large step forward in the detection of SPNs, and we expect that it will allow for the development of magnetic-field-based biosensors capable of detecting a single molecular binding event.

Super-paramagnetic nanoparticles (SPNs) are particles made of ferromagnetic material which, due to their small size, exhibit paramagnetic behavior with magnetic susceptibilities orders of magnitude larger than typical paramagnetic materials.[1] Magnetite ($Fe_3O_4$) SPNs can be readily functionalized for specific binding to a wide variety of molecules,[2–6] and are thus particularly useful for biological detection and imaging applications.[7–10] However, the detection of *single* SPNs in biologically compatible systems has remained an unsolved problem. Applications that would benefit greatly from this capability include highly sensitive assays for cancer,[11,12] HIV,[13] and non-acute-coronary-syndrome cardiac conditions.[14] In particular, digital immuno-assays rely on the detection of single bio-molecules to achieve ultra-high detection sensitivities.[15,16] In this work, we experimentally realize a room temperature platform capable of detecting single 20 nm diameter magnetite SPNs using wide-field optical imaging.

---

[*] mike.gould23@gmail.com

SPNs offer several advantages over conventional fluorescent tags. SPNs are detected magnetically, and thus the biological system under investigation can theoretically be completely isolated from optical fields, reducing undesired optical and thermal interactions. Furthermore, while the majority of biological samples exhibit fluorescence, they typically do not exhibit magnetism, potentially allowing for higher background contrast in SPN-based sensing schemes. Finally, SPN tags may allow for spatial manipulation of tagged molecules and the removal of unbound tags through the application of magnetic field gradients.[17,18]

Several SPN detection schemes have been demonstrated, such as giant magneto-resistive (GMR) sensors,[19,20] magnetic force microscopy (MFM),[21] superconducting quantum interference devices (SQUIDs),[22] micro-Hall sensors[23,24] and magnetic tunnel junctions (MTJs)[25]. Of the methods listed above, detection of single magnetic particles with diameters under 1 $\mu$m has only been demonstrated with SQUIDS and MFM. SQUIDs require operation at cryogenic temperatures while MFM makes use of a nano-mechanical scanning probe in contact with the sensing surface. Thus, neither of these platforms is ideal for biological integration.

Our platform uses negatively charged nitrogen-vacancy centers (NVs) in diamond to detect the magnetic field from SPNs. NVs are point defects in the diamond crystal lattice consisting of a substitutional nitrogen atom and a vacancy occupying nearest-neighbor lattice sites. The long electron spin coherence times of the NV ground state (>1 ms at room temperature),[26,27] combined with spin-dependent optical transitions, make NV centers attractive for highly sensitive magnetometry applications.[28,29] For example, single NV centers have been used to detect single nuclear spins within the diamond lattice.[30–32]

Ensembles of NVs can also be used for two-dimensional magnetic field imaging over larger regions, but with reduced sensitivity.[33–35] Using ensemble-based sensing, we are able to image disturbances in a uniform applied magnetic field due to the presence of SPNs.

NV centers can be viewed as a localized 2-electron system with the energy-level diagram shown in Figure 1(a). The ground state of the system is a spin-triplet, for which the $m_s = \pm 1$ spin states are degenerate under zero applied magnetic field, and split from the $m_s = 0$ spin state by an energy $E_{ss}$ due to spin-spin interactions.[36,37] Under an applied DC magnetic field, the Zeeman effect causes the $m_s = \pm 1$ ground states to split,[34,35] with respective energy shifts of $\Delta E = \pm g\mu_B \vec{B} \cdot \hat{z}$, where g is the electron g-factor, $\mu_B$ is the Bohr magneton, $\vec{B}$ is the applied DC magnetic field and $\hat{z}$ is a unit vector aligned with the NV symmetry axis. Thus by detecting $\Delta E$, the component of the applied field aligned with the NV axis can be determined.

Optical detection of $\Delta E$ is made possible by the presence of a decay path from the $m_s = \pm 1$ excited states, through two singlet states, to the $m_s = 0$ ground state.[37,38] This causes the $m_s = \pm 1$ states to emit less photoluminescence (PL) relative to the $m_s = 0$ state, and also allows for optical pumping to the $m_s = 0$ spin state. Under constant optical excitation, an RF magnetic field resonant with a ground state spin transition can be used to transfer population to an $m_s = \pm 1$ spin state, resulting in a decrease in detected PL. An example of an optically detected magnetic resonance (ODMR) curve is shown in Figure 1(b), in which collected PL intensity is measured as a function of RF frequency. A 3% dip is observed at 2.945 GHz, corresponding to the $m_s = 0$ to $m_s = -1$ spin transition for NV centers aligned to the applied field.

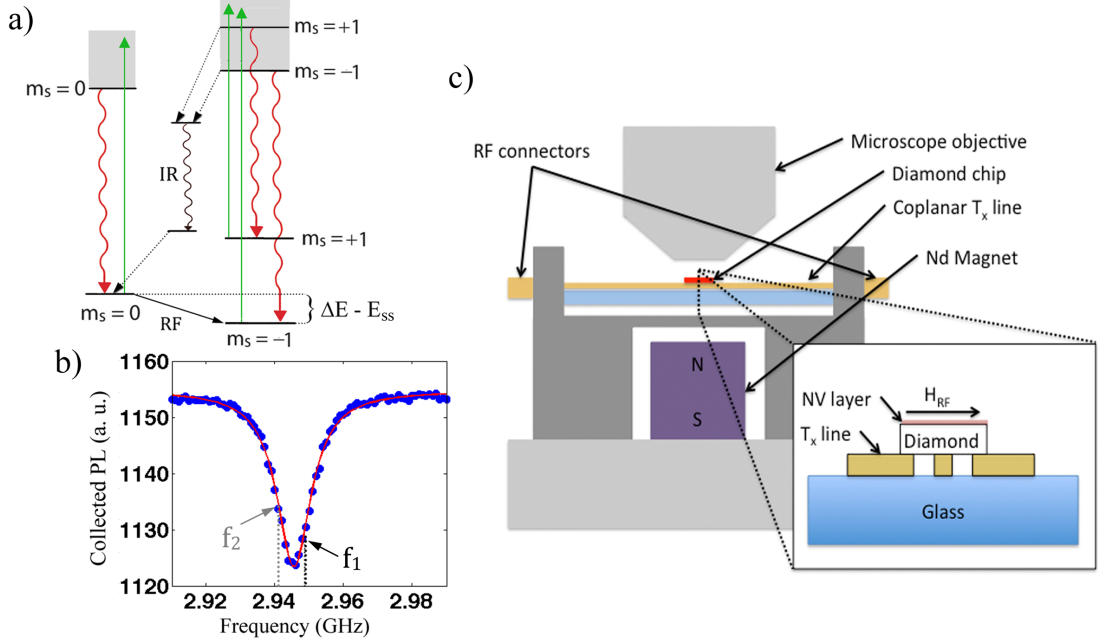

**Figure 1.** (a) Energy level diagram for an NV center. (b) Experimental ODMR curve taken with a ~200 mT magnetic field applied at the sensing surface. The bias point for the simple difference imaging scheme is shown in black and the second bias point for the 2-frequency scheme is shown in grey. Blue circles are experimental data points and the solid red line is a Lorentzian fit. (c) Schematic of setup below microscope objective showing the arrangement of the RF transmission line ($T_x$ line), Nd magnet, and diamond chip. Inset shows a zoomed-in cut through the diamond chip and $T_x$ line.

In this work, we perform imaging ODMR measurements on a 200 nm-thick, high-density sheet of NV centers near the {111} surface of a diamond chip. As depicted in Figure 1(c), the sensing chip sits on a co-planar transmission line used to apply an RF magnetic field in the plane of the chip surface. A neodymium (Nd) magnet sits below, applying a DC magnetic field of approximately 200 mT at the surface, aligned to the surface normal and thus to one of the four distinguishable <111> NV orientations. The chip is imaged using standard wide-field fluorescence microscopy, with a 10 $\mu$m diameter excitation area and a resolution of approximately 500 nm.

Using this setup, the spatial distribution of the component of the magnetic field normal to the chip surface is imaged. We use a simple difference scheme utilizing only NVs aligned with the applied DC field. Images taken with an RF magnetic field applied to the sensing surface are subtracted from images with no applied RF. The frequency of the RF field is set just below the steepest part of the ODMR curve ($f_1$ in Figure 1(b)), maintaining a high small-field sensitivity while increasing the signal from the high field in proximity to SPNs.

In order to characterize the system's SPN detection capability, magnetite SPNs were deposited on the sensing surface in isolated single particles and small groups. The SPNs were prepared according to a previously reported method.[39–42] Particle distributions were obtained by drying colloidal suspensions on a lithographically defined pattern on the sensor surface, resulting in a grid pattern with small groups of particles at each grid point. Figures 2(a-b) show scanning electron microscope (SEM) images of a resulting pattern of particle groups.

Figure 2(c) shows a magneto-optical image of the area shown in Figure 2(a), obtained using the difference-imaging scheme described above. The total exposure time for each RF field state was 10 s. The magnetic field limit-of-detection (LOD) is 5.0 $\mu$T, and does not improve with increased integration time. As shown, this imaging scheme allows for clear detection of groups containing 3 or more particles. However, the group of 2 particles exhibits a signal very close to the noise floor. There is no detectable signal from the single particle. In comparison, no particle group is detected in the corresponding PL image shown in Figure 2(d), as expected.

The experimental LOD is in good agreement with the expected value of 5.4 $\mu$T, based on experimental ODMR spectra from the chip, and the optical resolution of the system. However, the observed lack of improvement in the LOD with longer integration time suggests that field detection is limited by some source of time-invariant spatial fluctuation in the ODMR signal. In order to identify the source of the background spatial variation, we took simultaneous ODMR spectra for 25 different spots within a 2.5 $\mu$m x 2.5 $\mu$m area devoid of any magnetic material. The average depth of the ODMR dips for this data was 3%, with a standard deviation of 0.04%. For the simple difference scheme described above, this corresponds to an LOD of 10.6 $\mu$T, which is larger than the measured LOD. This suggests variation in the depth of the ODMR dip as the limiting factor in our system. It is likely that the variation is not constant over the chip, and in the regions where particle detection was performed, the ODMR characteristics were slightly more uniform, limiting the system to 5 $\mu$T rather than 10.6 $\mu$T. Importantly, the spatial distribution of the ODMR depth remained constant over several runs, indicating that the variation was physical and not a result of noise from the curve-fitting algorithm.

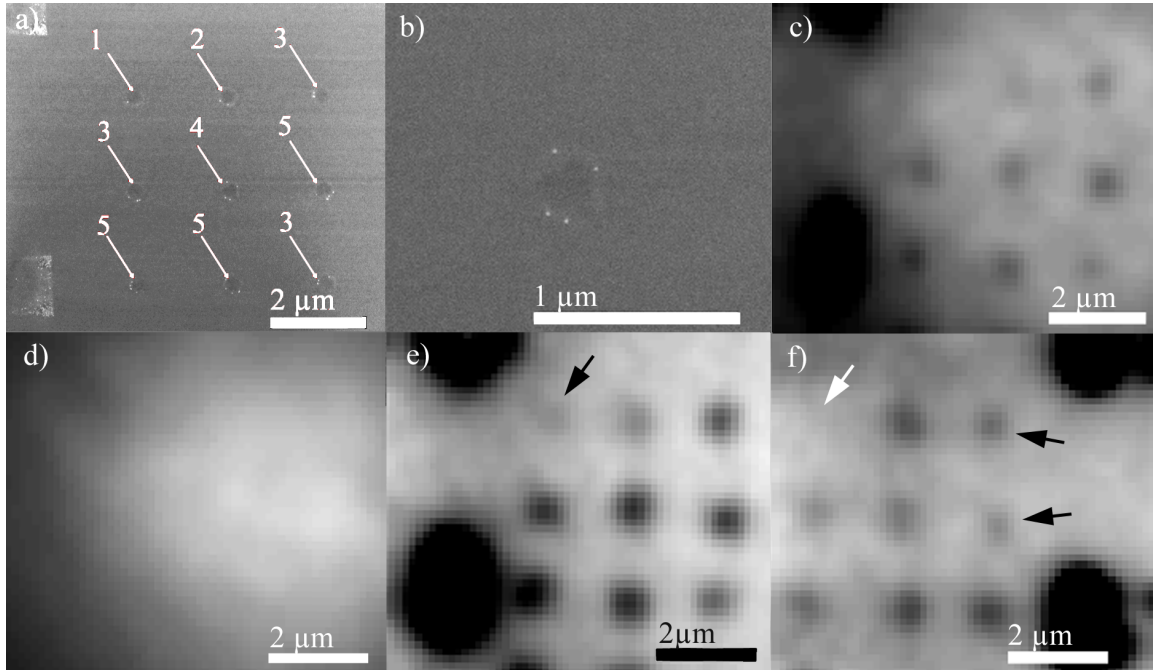

**Figure 2.** (a) SEM image of small groups of SPNs arranged in a grid pattern, with particle number noted for each group. (b) Digital zoom on the central grid point of (a), showing a group of 4 particles. (c) A magneto-optical image of area shown in (a). (d) PL image of area shown in (a). (e) Magneto-optical image of area shown in (a), using 2-frequency imaging, with the black arrow indicating the single particle. (f) Magneto-optical image of second area using 2-frequency scheme, with black arrows indicating the single particles and the white arrow indicating an empty grid spot.

The effect of this variation on the system's LOD can be mitigated by taking the difference between images taken at two different frequencies, on either side of the background ODMR resonance. This new scheme accentuates the effect of changes that are anti-symmetric about the resonance, such as resonance shifts, with respect to changes that are symmetric about the resonance, such as depth variations. For particle detection, we find that our signal-to-noise ratio (SNR) is optimized when the frequencies are chosen to be approximately 1 MHz below the steepest points on the ODMR curve, as shown in Figure 1(b). Once again, this can be understood as a slight move away from the highest small-field sensitivity in order to obtain more signal from the high field strengths in close proximity to the SPNs. Using this imaging scheme, the LOD is improved to 2.3 $\mu$T for a 10 s integration time. Figure 2(e) shows the resulting magneto-optical image of the same area as shown in Figures 2(a-d). Figure 2(f) shows a magneto-optical image of a second area, also containing single. In both images, single particles, indicated by black arrows, are clearly visible. The upper left grid spot in Figure 2(f), indicated with a white arrow, contains no SPNs and there is correspondingly no detected signal.

The detection of single particles and small groups using the 2-frequency scheme is summarized in Figure 3, where the SNR for each group is plotted as a function of particle count. Importantly, all groups and single particles were detected with SNR values greater than 1.5. Furthermore, this data suggests particle number can be determined to within ±1 particle for group sizes < 5. The variation in signal strength for a given group size is likely due to variations in particle size. In fact, the measured size variation for the particles used in this work corresponds to a variation in magnetic moment of a factor of 5.

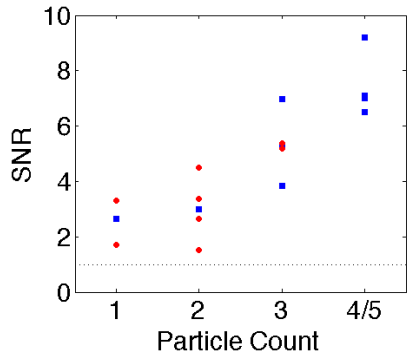

**Figure 3.** Plot of SNR as a function of particle count for all 17 groups in both imaging areas (Figures 2(e) and 2(f)). Blue squares correspond to groups in the first imaging area while red circles represent groups in the second. Dashed line is noise floor (SNR = 1).

We suspect that the 2.3 $\mu$T LOD of the 2-frequency scheme is still limited by spatial variations of the ODMR characteristics across the chip. For example, local shifts in the frequency of the spin transitions due to strain variation[36,37,43] cannot be removed by the two-frequency difference scheme. We expect to be able to mitigate this effect and further improve the limit-of-detection of the system by carefully measuring the ODMR spectrum of each pixel in the imaging area prior to any sensing experiments, allowing for subsequent computational image correction. Other possible sources of spatial noise are systematic spatial and/or thermal shifts between RF states, which could be reduced by mechanical and thermal isolation of RF components from the rest of the system. Additional paths to improved detection include using samples with longer spin-dephasing times, the use of thinner NV layers obtained using He$^+$ implantation[44,45] or by direct incorporation of NV centers during epitaxial diamond growth,[46] and improved PL collection efficiency.[47] Ultimate sensitivity will result from an optimized NV density balancing spin coherence time and PL brightness,[28] an NV layer thickness optimized for the particles to be detected, and collection efficiency into the first objective much higher than the estimated 3% obtained on the current setup.

In summary, we have successfully demonstrated room temperature single particle detection of 20 nm magnetite SPNs using wide-field optical microscopy. The significant advantage of this system over others capable of single SPN detection is its relative simplicity; detection does not require cryogenic temperatures or nano-mechanical components, leaving it open to integration with biological and other nano-scale surface-based experiments. This marks a significant step forward in the detection of SPNs, and we expect that it will allow for the development of bio-sensors capable of detecting a single molecular surface binding event. Continued improvements in sensitivity will allow for further applications in biological detection and imaging, as well as broader applications in the study of nano-scale magnetic systems.

The authors would like to acknowledge support from the Washington Nanofabrication Facility, and would like to thank Richard Bojko for electron beam lithography support as well as R. M. Ferguson for providing nanoparticle samples. Financial support for this work was provided by the University of Washington Royalty Research Fund award A79342. M. Gould received financial support from the Natural Science and Engineering Research Coucil of Canada through a Post-Graduate Scholarship.

1. Bean, C. & Livingston, J. Superparamagnetism. *J. Appl. Phys.* **30,** S120 (1959).

2. Lu, H. *et al.* Synthesis and characterization of multi-functional nanoparticles possessing magnetic, up-conversion fluorescence and bio-affinity properties. *J. Mater. Chem.* **14,** 1336 (2004).

3. Veiseh, O. *et al.* Specific targeting of brain tumors with an optical/magnetic resonance imaging nanoprobe across the blood-brain barrier. *Cancer Res.* **69,** 6200–7 (2009).

4.  Sun, C. *et al*. PEG-mediated synthesis of highly dispersive multifunctional superparamagnetic nanoparticles: their physicochemical properties and function in vivo. *ACS Nano* **4,** 2402–10 (2010).

5.  Jin, Y., Jia, C., Huang, S.-W., O'Donnell, M. & Gao, X. Multifunctional nanoparticles as coupled contrast agents. *Nat. Commun.* **1,** 41 (2010).

6.  Gao, J., Gu, H. & Xu, B. Multifunctional magnetic nanoparticles: design, synthesis, and biomedical applications. *Acc. Chem. Res.* **42,** 1097–107 (2009).

7.  Brzeska, M. *et al*. Detection and manipulation of biomolecules by magnetic carriers. *J. Biotechnol.* **112,** 25–33 (2004).

8.  Chemla, Y. R. *et al*. Ultrasensitive magnetic biosensor for homogeneous immunoassay. *Proc. Natl. Acad. Sci. U. S. A.* **97,** 14268–72 (2000).

9.  Haun, J. & Yoon, T. Magnetic nanoparticle biosensors. *Wiley Interdiscip. Rev. Nanomedicine Nanobiotechnology* **2,** 291–304 (2010).

10. Krishnan, K. Biomedical Nanomagnetics : A Spin Through Possibilities in Imaging, Diagnostics, and Therapy. *Magn. IEEE Trans*. **46,** 2523–2558 (2010).

11. Wulfkuhle, J. D., Liotta, L. A. & Petricoin, E. F. Proteomic applications for the early detection of cancer. *Nat. Rev. Cancer* **3,** 267–75 (2003).

12. Liu, X. *et al*. A one-step homogeneous immunoassay for cancer biomarker detection using gold nanoparticle probes coupled with dynamic light scattering. *J. Am. Chem. Soc*. **130,** 2780–2 (2008).

13. Miedouge, M., Grèze, M., Bailly, A. & Izopet, J. Analytical sensitivity of four HIV combined antigen/antibody assays using the p24 WHO standard. *J. Clin. Virol*. **50,** 57–60 (2011).

14. De Lemos, J. A. Increasingly sensitive assays for cardiac troponins: a review. *JAMA* **309,** 2262–9 (2013).

15. Wilson, D. H. *et al*. Fifth-generation digital immunoassay for prostate-specific antigen by single molecule array technology. *Clin. Chem*. **57,** 1712–21 (2011).

16. Chang, L. *et al*. Simple diffusion-constrained immunoassay for p24 protein with the sensitivity of nucleic acid amplification for detecting acute HIV infection. *J. Virol. Methods* **188,** 153–60 (2013).

17. Kim, K. S. & Park, J.-K. Magnetic force-based multiplexed immunoassay using superparamagnetic nanoparticles in microfluidic channel. *Lab Chip* **5,** 657–64 (2005).

18. Hoffmann, C. *et al*. Spatiotemporal control of microtubule nucleation and assembly using magnetic nanoparticles. *Nat. Nanotechnol.* **8,** 199–205 (2013).

19. Li, G. *et al*. Detection of single micron-sized magnetic bead and magnetic nanoparticles using spin valve sensors for biological applications. *J. Appl. Phys.* **93,** 7557 (2003).

20. Gaster, R. S. *et al*. Quantification of protein interactions and solution transport using high-density GMR sensor arrays. *Nat. Nanotechnol.* **6,** 314–320 (2011).

21. Nocera, T., Chen, J., Murray, C. & Agarwal, G. Magnetic anisotropy considerations in magnetic force microscopy studies of single superparamagnetic nanoparticles. *Nanotechnology* **23,** 495704 (2012).

22. Thirion, C., Wernsdorfer, W. & Mailly, D. Switching of magnetization by nonlinear resonance studied in single nanoparticles. *Nat. Mater.* **2,** 524–527 (2003).

23. Gabureac, M., Bernau, L., Boero, G. & Utke, I. Single superparamagnetic bead detection and direct tracing of bead position using novel nanocomposite Hall sensors. *IEEE Trans. Nanotechnol.* **12,** 668–673 (2013).

24. Mihajlovic, G. & Xiong, P. Detection of single magnetic bead for biological applications using an InAs quantum-well micro-Hall sensor. *Appl. Phys. Lett.* **87,** 112502 (2005).

25. Shen, W., Liu, X., Mazumdar, D. & Xiao, G. In situ detection of single micron-sized magnetic beads using magnetic tunnel junction sensors. *Appl. Phys. Lett.* **86,** 253901 (2005).

26. Bar-Gill, N., Pham, L. M., Jarmola, a, Budker, D. & Walsworth, R. L. Solid-state electronic spin coherence time approaching one second. *Nat. Commun.* **4,** 1743 (2013).

27. Balasubramanian, G. *et al*. Ultralong spin coherence time in isotopically engineered diamond. *Nat. Mater.* **8,** 383–7 (2009).

28. Taylor, J., Cappellaro, P. & Childress, L. High-sensitivity diamond magnetometer with nanoscale resolution. *Nat. Phys.* **4,** 810–816 (2008).

29. Degen, C. L. Scanning magnetic field microscope with a diamond single-spin sensor. *Appl. Phys. Lett.* **92,** 243111 (2008).

30. Zhao, N., Honert, J., Schmid, B. & Klas, M. Sensing single remote nuclear spins. *Nat. Nanotechnol.* **7,** 657–662 (2012).

31. Neumann, P. *et al*. Single-shot readout of a single nuclear spin. *Science* **329,** 542–4 (2010).

32. Robledo, L. *et al*. High-fidelity projective read-out of a solid-state spin quantum register. *Nature* **477,** 574–8 (2011).

33. Steinert, S. *et al*. High sensitivity magnetic imaging using an array of spins in diamond. *Rev. Sci. Instrum.* **81,** 043705 (2010).

34. Pham, L. & Sage, D. Le. Magnetic field imaging with nitrogen-vacancy ensembles. *New J. Phys.* **13,** 045021 (2011).

35. Maertz, B. J., Wijnheijmer, A. P., Fuchs, G. D., Nowakowski, M. E. & Awschalom, D. D. Vector magnetic field microscopy using nitrogen vacancy centers in diamond. *Appl. Phys. Lett.* **96,** 092504 (2010).

36. Manson, N., Harrison, J. & Sellars, M. Nitrogen-vacancy center in diamond: Model of the electronic structure and associated dynamics. *Phys. Rev. B* **74,** 104303 (2006).

37. Lenef, A. & Rand, S. Electronic structure of the N-V center in diamond: Theory. *Phys. Rev. B* **53,** (1996).

38. Acosta, V., Jarmola, A., Bauch, E. & Budker, D. Optical properties of the nitrogen-vacancy singlet levels in diamond. *Phys. Rev. B* **82,** 201202 (2010).

39. Arami, H. & Ferguson, R. Size-dependent ferrohydrodynamic relaxometry of magnetic particle imaging tracers in different environments. *Med. Phys.* **40,** 071904 (2013).

40. Ferguson, R. M., Khandhar, A. P. & Krishnan, K. M. Tracer design for magnetic particle imaging (invited). *J. Appl. Phys.* **111,** 7B318–7B3185 (2012).

41. Khandhar, A. P., Ferguson, R. M., Simon, J. A. & Krishnan, K. M. Tailored magnetic nanoparticles for optimizing magnetic fluid hyperthermia. *J. Biomed. Mater. Res. A* **100,** 728–37 (2012).

42. Khandhar, A. P., Ferguson, R. M., Arami, H. & Krishnan, K. M. Monodisperse magnetite nanoparticle tracers for in vivo magnetic particle imaging. *Biomaterials* **34,** 3837–45 (2013).

43. Olivero, P. *et al*. Splitting of photoluminescent emission from nitrogen–vacancy centers in diamond induced by ion-damage-induced stress. *New J. Phys.* **15,** 043027 (2013).

44. Huang, Z. *et al*. Diamond nitrogen-vacancy centers created by scanning focused helium ion beam and annealing. *Appl. Phys. Lett*. **103,** 081906 (2013).

45. McCloskey, D. *et al*. Helium ion microscope generated nitrogen-vacancy centres in type Ib diamond. *Appl. Phys. Lett*. **104,** 031109 (2014).

46. Ishikawa, T. *et al*. Optical and spin coherence properties of nitrogen-vacancy centers placed in a 100 nm thick isotopically purified diamond layer. *Nano Lett*. **12,** 2083–7 (2012).

47. Yeung, T. K., Le Sage, D., Pham, L. M., Stanwix, P. L. & Walsworth, R. L. Anti-reflection coating for nitrogen-vacancy optical measurements in diamond. *Appl. Phys. Lett*. **100,** 251111 (2012).